\documentclass[prd,floats,floatfix,showpacs,nofootinbib]{revtex4}
\usepackage{graphicx}
\usepackage{dcolumn}
\usepackage{bm}
\usepackage{graphics}
\usepackage{slashed}
\usepackage{amssymb}
\usepackage{natbib}
\usepackage{amsmath}
\usepackage{amsthm}
\usepackage{url}
\usepackage{color}
\newcommand{\fracc}[2]{\frac{\textstyle{#1}}{\textstyle{#2}}}

\begin{document}

\title{Friedmann-like solutions with a non-vanishing Weyl tensor}

\author{E. Bittencourt}\email{eduhsb@cbpf.br}
\author{J. M. Salim}\email{jsalim@cbpf.br}

\affiliation{Instituto de Cosmologia Relatividade e Astrofisica ICRA -
CBPF\\ Rua Doutor Xavier Sigaud, 150, CEP 22290-180, Rio de Janeiro,
Brazil}

\date{\today}

\begin{abstract}
We have solved the Einstein equations of general relativity for a class of metrics with constant spatial curvature and found a non-vanishing Weyl tensor in the presence of an energy-momentum tensor with an anisotropic pressure component. The time evolution of the spacetime is guided by the usual Friedmann equations and the constraints on the hypersurface comprise a separated system of equations that can be independently solved. Contrary to the apparent behavior induced by some choices of coordinates, the metric we have obtained is completely regular everywhere and is free of singularities (except the well-known Friedmann singularity at $t=0$). The physical features of this solution are elucidated by using the Quasi-Maxwellian equations (a set of third order differential equations describing the dynamics of the gravitational field in terms of the Weyl tensor). The motion of test particles is also analyzed in order to confirm the maximal extension of the manifold under consideration. These results indicate that the anisotropic pressure could mimic dark matter effects on certain geodesic congruences keeping the cosmic flow unchanged.
\end{abstract}

\maketitle

\section{Introduction}
One of the fundamental postulates of general relativity (GR) is the equivalence principle. This postulate states that, due to the universality of the gravitational field, it can be set to zero locally (at most along a geodesic) by a change of coordinates. In other words, the gravitational effects upon a fluid containing up to first derivatives of the newtonian potential, which corresponds to first derivatives of the metric tensor, can be eliminated by a change of reference frame. The equivalence principle plus the principle of general covariance provide the conservation law of the energy-momentum tensor of the fluid with respect to the spacetime metric.

The question we address in this paper, by providing a concrete example in the cosmological scenario, is that gravitational effects which manifest themselves only from second derivatives of the metric and, therefore, cannot be cancelled through the equivalence principle, can be associated to viscous process in fluid mechanics. Such effect can be understood as being consequence of the non-negligible gravitational interaction at the microscopic level between the constituting particles of the fluid resulting in dissipative terms and producing non-local effects of purely gravitational origin that are not geometrized in GR.

To clarify our exposition, we present our proposal from the point of view of two distinct approaches of GR: one is given by the Einstein formulation \cite{einstein} and the other one is given by the Quasi-Maxwellian (QM) equations \cite{jek}. Once the QM equations involve a higher order of differentiability than GR, according to Lichnerowicz's theorem \cite{lich}, these formulations are equivalent only if appropriate initial conditions are used. It is clear that the theorem constrains the solutions of the QM equations in order that both formulations surely give the same results. The critical point of this restriction appears when we want to determine the initial data in terms of observable quantities aiming to guarantee that the solution we shall obtain has something to do with the empirical ingredients we started with. It is well known that the observables in GR can only be determined through geodesic deviation and that they affect the curvature tensor. This tensor has 20 independent components that can be expressed in terms of the Ricci tensor, the scalar curvature and the Weyl tensor. In the usual formulation of the Einstein equations only the Ricci tensor and the scalar curvature are present. They are written in terms of derivatives of $g_{ij}$ and nothing can be inferred about the Weyl tensor before obtaining the solution of the metric. The main issue is that many empirical data are directly represented by the Weyl tensor, for example, the tidal forces or non-vanishing gravitational fields in the absence of matter. In other words, in Einstein's formulation, it is not possible to look for a complete solution of GR corresponding to a determined Weyl tensor specified by a given set of empirical data. Such information cannot be expressed in terms of the initial values of the variables of the Einstein equations.

All these difficulties do not appear in the QM formulation of gravity. In this approach, the variables of the theory represent directly the empirical data and the Einstein equations are used (as a first integral) to relate the Weyl tensor to the energy-momentum tensor through differential equations involving both. In particular, in this paper, we present how this relation takes place when we analyze the standard cosmological model (SCM).

The SCM described by FLRW metrics (Friedmann-Lema\^itre-Robertson-Walker) experiences difficulties in which exotic components of matter and energy are introduced in an attempt to explain, for instance, the apparent accelerated expansion of the universe (dark energy), the galaxy rotation curves (dark matter) or the initial singularity in the far past of its history. Some authors claim that the main problem lies on the huge simplification of the geometry adopted and hence they suggest modifications of spacetime symmetries as in the case of inhomogeneous models \cite{ltb}, in particular indicating the differences between average processes \cite{wiltshire}, or modifications in the coupling between matter and geometry as displayed by some bouncing cosmologies \cite{mario_sant}, or even non-conventional proposals as \cite{hoyle}, which change completely our understanding upon the spacetime itself.

Notwithstanding, our proposal is simpler. We have shown that even metrics very similar to the Friedmann one (which we shall specify afterwards) admit a more general solution when we introduce an anisotropic pressure term $\pi_{\mu\nu}$ in the Einstein equations. As mentioned by \cite{ltb}, such term has been considered in the establishment of some cosmological models, but two simple reasonings make it an undesirable term in cosmology: first, it violates all the symmetries of the FLRW metrics; second, in the case of shear-free geometries (e.g. the Friedmann metric) there is no other traceless symmetric tensor phenomenologically linked to $\pi_{\mu\nu}$. Both arguments are not completely true and we shall see that this term can be very important in the transition from an inhomogeneous model to the FLRW models, because the dynamics of the solution is given by Friedmann equations not only in asymptotic regimes \cite{ltb}.

Summarizing, in the present paper we show that if one considers a larger class of geometries than the FLRW ones, restricted by the imposition ${\cal M}^4(g_{\mu\nu})={\cal M}^3(h_{\mu\nu})\bigotimes\mathbb{R}$, where ${\cal M}^3(h_{\mu\nu})$ possesses constant scalar curvature, and an energy-momentum tensor representing a simple fluid with an anisotropic pressure, the Einstein equations allow a more general solution which has the usual time evolution given by the Friedmann equations and a 3-space given by the Schwarzschild lattice, for vanishing spatial curvature. However, some geometrical properties remain obscure in the Einstein formalism. On the other hand, using the QM equations we immediately obtain a non-zero Weyl tensor given in terms of the anisotropic pressure. Therefore, assuming the Lichnerowicz theorem, we are led to affirm that both formulations should be considered as complementary from the point of view of the determination of initial conditions in GR\footnote{Some attempts have been made in order to show that the class of solutions of the QM equations is larger than the Einstein's one, as it can be seen in Ref. \cite{nov_trieste}.}. In the next section we revisit the Friedmann solution in order to do a self-consistent exhibition of our results.

\section{Friedmann solution revisited}

The observational data indicates that the Friedmann geometry is more convenient to describe our universe, mainly because of its homogeneity and isotropy at large scales. These symmetries suggest that the only possible fluid satisfying these properties is a perfect fluid with energy density $\rho$ and isotropic pressure $p$. However, as we mentioned before, this is not a paradigm. In this paper, we consider an imperfect fluid as source for the gravitational field and some special symmetries of spacetime are still preserved.

To make a self-consistent exposition of our results, we present here a brief derivation of the Friedmann model. Thus, let us start considering the infinitesimal line element given by

\begin{equation}
\label{fried}
ds^2=dt^2-a^2(t)[d\chi^2+\sigma^2(\chi)d\Omega^2],
\end{equation}
where $t$ represents the cosmic time, $a(t)$ is the scale factor and $\sigma(\chi)$ is an arbitrary function of the spatial coordinate $\chi$. We call it a Friedmann-like metric.

A straightforward calculation gives the following scalar curvature

\begin{equation}
\label{4_ricci}
R=6\fracc{\ddot a}{a} + 6\fracc{\dot a^2}{a^2} - \fracc{2}{a^2}\left(2\fracc{\sigma''}{\sigma} + \fracc{\sigma'^2}{\sigma^2} - \fracc{1}{\sigma^2}\right),
\end{equation}
where dot ($\,\dot{} \,$) means time derivative and prime ($\,'\,$) means derivative w.r.t. $\chi$. The spatial curvature $^{(3)}R$ of the hypersurface defined by $t\equiv const.$ is

\begin{equation}
\label{3_ricci}
^{(3)}R=-2\left(2\fracc{\sigma''}{\sigma}+\fracc{\sigma'^2}{\sigma^2}-\fracc{1}{\sigma^2}\right).
\end{equation}
Assuming that $^{(3)}R$ has the same value everywhere, we set
$$^{(3)}R\equiv6\epsilon,$$
where $\epsilon$ is a constant. Therefore, the scalar curvature becomes

\begin{equation}
\label{4_ricci_simp}
R=6\left(\fracc{\ddot a}{a}+\fracc{\dot a^2}{a^2}+\fracc{\epsilon}{a^2}\right).
\end{equation}
This equation shows that the scalar curvature of the spacetime depends only on time.

The energy-momentum distribution is described by a perfect fluid with energy density $\rho$, isotropic pressure $p$ and comoving four-velocity $V^{\mu}=\delta^{\mu}_0$, namely
$$T_{\mu\nu}=(\rho+p)V_{\mu}V_{\nu}-pg_{\mu\nu}.$$
In general, one assumes the existence of an equation of state such that $p=\lambda\,\rho$, where $\lambda$ is a constant. In this way, we can calculate the non-trivial components of the Einstein equations $G^{\mu}{}_{\nu}=-T^{\mu}{}_{\nu}$ (the Einstein constant is set to $1$), which are explicitly given by

\begin{subequations}
\label{eins_tens}
\begin{eqnarray}
&&3\fracc{\dot a^2}{a^2}+3\fracc{\epsilon}{a^2}=\rho,\label{eins_tens0}\\[2ex]
&&2\fracc{\ddot a}{a} + \fracc{\dot a^2}{a^2} + \fracc{3\epsilon}{a^2} + \fracc{2}{a^2}\fracc{\sigma''}{\sigma} = -p,\label{eins_tens1}\\[2ex]
&&2\fracc{\ddot a}{a} + \fracc{\dot a^2}{a^2} - \fracc{1}{a^2}\fracc{\sigma''}{\sigma} = -p.\label{eins_tens2}
\end{eqnarray}
\end{subequations}

Subtracting Eq. (\ref{eins_tens1}) from Eq. (\ref{eins_tens2}) yields

\begin{equation}
\label{sigma_eq}
\fracc{\sigma''}{\sigma}+\epsilon=0.
\end{equation}
Alternatively, we can combine Eqs.\ (\ref{sigma_eq}) and (\ref{3_ricci}) to obtain the equation

\begin{equation}
\label{sigma_eq2}
\fracc{\sigma''}{\sigma}-\fracc{\sigma'^2}{\sigma^2}+\fracc{1}{\sigma^2}=0.
\end{equation}
One can see that each value of the spatial curvature corresponds to a single curve in the space of solutions and it obviously happens because Eqs.\ (\ref{sigma_eq}) and (\ref{sigma_eq2}) must be satisfied simultaneously. Note that solutions with the same sign for $\epsilon$ are topologically equivalent. It means that $sign(\epsilon)$ is enough to characterize all the solutions of Eq.\ (\ref{sigma_eq}). Therefore, Eqs.\ (\ref{sigma_eq}) and (\ref{sigma_eq2}) have only three relevant solutions, namely

\begin{equation}
\label{sigma_eq_sol}
\left\{\begin{array}{lcl}
\epsilon=0&\Longrightarrow&\sigma=\chi,\\[2ex]
\epsilon=1&\Longrightarrow&\sigma=\sin \chi,\\[2ex]
\epsilon=-1&\Longrightarrow&\sigma=\sinh \chi.
\end{array}\right.
\end{equation}

We can make a coordinate transformation given by $r=\sigma(\chi)$ to explicitly exhibit the metric in spherically symmetric coordinates. The line element thus becomes

\begin{equation}
\label{fried_r}
ds^2=dt^2-a^2(t)\left(\fracc{dr^2}{1-\epsilon r^2}+r^2d\Omega^2\right).
\end{equation}
Note that this metric is conformally equivalent to Minkowski, de Sitter and anti-de Sitter spacetimes for $\epsilon$ equal to $0$, $-1$ and $1$, respectively. Using this coordinate system, the time evolution of the scale factor is given by

\begin{subequations}
\label{gr_fried_solved}
\begin{eqnarray}
&&H^2=\fracc{\rho}{3}-\fracc{\epsilon}{a^2},\label{gr_fried_solved1}\\[2ex]
&&\fracc{\ddot a}{a}=-\fracc{1}{6}(1+3\lambda)\rho,\label{gr_fried_solved2}
\end{eqnarray}
\end{subequations}
where $H\equiv\dot a/a$ is the Hubble parameter. The system of equations\ (\ref{gr_fried_solved}) corresponds to the well-known Friedmann equations. A standard analysis of this model can be found in \cite{mukhanov} and references therein.

In terms of the Quasi-Maxwellian formalism (see Appendix\ [\ref{appendix}]), the Friedmann equations are equivalent to

\begin{subequations}
\label{gr_fried}
\begin{eqnarray}
&&\dot\theta+\fracc{\theta^2}{3}=-\fracc{1}{2}(1+3\lambda)\rho,\label{gr_fried1}\\[2ex]
&&\dot\rho+(1+\lambda)\rho\,\theta=0,\label{gr_fried2}
\end{eqnarray}
\end{subequations}
which are the Raychaudhuri equation and the continuity equation, respectively. In this case, we observe that the QM equations are redundant to the Einstein equations, because they do not give any additional information about the system, although they express the Friedmann equations directly in terms of measurable physical quantities. This will not be the case when we introduce the anisotropic pressure, as we shall see in the next sections.

\section{Friedmann model in the presence of an anisotropic pressure}

In this section, we basically add to the Einstein equations an extra term on the right hand side corresponding to the presence of the anisotropic pressure $\pi_{\mu\nu}$. In the formalism of fluid mechanics \cite{landau}, this term describes all processes involving viscosity and, consequently, energy dissipation. Before the eighties, this term was commonly used in GR \cite{bel_kha}. Notwithstanding, with the advent of the SCM, we usually do not consider sources for the gravitational field including such term anymore. At most, we can find some authors dealing with corrections to the isotropic pressure producing a dissipative fluid \cite{salim,nov_olival,bel_kha,zimdahl,pavon}. Concerning only the thermodynamics, some references analyze phase transitions produced by the gravitational field in the presence of the anisotropic pressure \cite{nov_duque,nov_duque2}. However, none of these examples take into account the anisotropic pressure on the right hand side of the Einstein equations and try to solve them.

To do so, we propose that the most general source for a Friedmann-like geometry is represented by\footnote{We cannot add the heat flux $q^{\mu}$ because it breaks the isotropy of the spacetime.}
$$T_{\mu\nu}=(\rho+p)V_{\mu}V_{\nu}-p\,g_{\mu\nu}+\pi_{\mu\nu}.$$
Starting from the line element given by Eq.\ (\ref{fried}), the non-trivial components of the Einstein equations are

\begin{subequations}
\label{eins_tens_pi}
\begin{eqnarray}
&&3\fracc{\dot a^2}{a^2}+3\fracc{\epsilon}{a^2}=\rho,\label{eins_tens0_pi}\\[2ex]
&&2\fracc{\ddot a}{a} + \fracc{\dot a^2}{a^2} + \fracc{3\epsilon}{a^2} + \fracc{2}{a^2}\fracc{\sigma''}{\sigma} = -p+\pi^1{}_1,\label{eins_tens1_pi}\\[2ex]
&&2\fracc{\ddot a}{a} + \fracc{\dot a^2}{a^2} - \fracc{1}{a^2}\fracc{\sigma''}{\sigma} = -p+\pi^2{}_2.\label{eins_tens2_pi}
\end{eqnarray}
\end{subequations}
The off diagonal components are identically zero.

\begin{figure}[!htb]
\centering
\includegraphics[width=6cm,height=8cm,angle=-90]{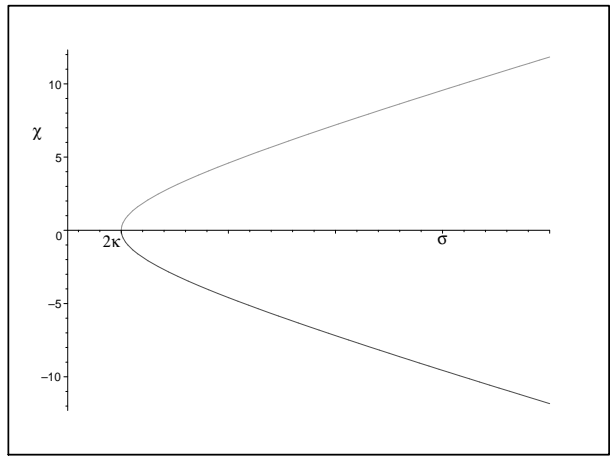}
\caption{Relation between $\sigma$ and $\chi$, for $\epsilon=0$. Note that $\sigma$ has a minimal value in $2k$.}
\label{fig1}
\end{figure}

Subtracting Eq. (\ref{eins_tens1_pi}) from Eq. (\ref{eins_tens2_pi}) and using the traceless condition $\pi^{\mu}{}_{\mu}=0$, we obtain

\begin{equation}
\label{sigma_eq_pi}
\fracc{\sigma''}{\sigma}+\epsilon=\fracc{1}{2}f(\sigma).
\end{equation}
The Einstein equations admit an anisotropic pressure term only if $\pi^{\mu}_{\nu}$ can be written as an arbitrary function of $\sigma$ times a time dependent, namely
$$\pi^{2}{}_{2} = \pi^{3}{}_{3}, \hspace{1cm} \pi^{1}{}_{1} = -2\pi^{2}{}_{2}, \hspace{1cm} \pi^{1}{}_{1}=\fracc{f(\sigma)}{a^2},$$
where the factor $1/a^2$ in the expression of $\pi^1{}_1$ is introduced for consistency reasons. Alternatively, we can combine Eqs.\ (\ref{sigma_eq_pi}) and (\ref{3_ricci}) to obtain
\begin{equation}
\label{sigma_eq2_pi}
\fracc{\sigma''}{\sigma}-\fracc{\sigma'^2}{\sigma^2}+\fracc{1}{\sigma^2}=\fracc{3}{2}f.
\end{equation}
However, the solutions of Eqs.\ (\ref{3_ricci}) and\ (\ref{sigma_eq2_pi}) are the same only if we impose
$$f=\fracc{2k}{\sigma^3},$$
where $k$ is an integration constant\footnote{In this paper we concentrate only on the case $k>0$ motivated by the geodesic analysis in Sec.\ [\ref{test_part}].}. This condition emerges from the compatibility relation of the first integrals of both equations, which is

\begin{equation}
\label{dif_sigma}
\sigma'=\pm\sqrt{1-\epsilon \sigma^2-\fracc{2k}{\sigma}}.
\end{equation}
For $k\neq0$, the integration can be done analytically only for $\epsilon=0$, resulting in

\begin{equation}
\label{sol_sigma}
\chi = \pm\left[ \sqrt{\sigma^2-2k\sigma} + k\ln\left(\sigma-k+\sqrt{\sigma^2-2k\sigma} \right)\right].
\end{equation}
This expression gives $\sigma$ implicitly in terms of $\chi$. Moreover, it covers the whole manifold, because the range of each spatial coordinate is maximally extended. From Fig.\ (\ref{fig1}) we clearly see that $\sigma$ has a minimal value at $2k$. It means that the region $\sigma<2k$ is excluded from this manifold. Eq.\ (\ref{dif_sigma}) corroborates this statement since $\sigma'$ becomes complex in this region. This result is very important and we shall use it to analyze the geodesic motion and demonstrate the completeness and smoothness of our solution, in this specific case $\epsilon=0$, in Sec.\ [VI]\footnote{We also have calculated all Debever invariants and verified that they are regular at $\sigma=2k$, although we have not displayed them here.}.

In the general case, depending on the value of $\epsilon$ and $k$, the polynomial inside the square root in Eq.\ (\ref{dif_sigma}) has different numbers of roots. Thus, the analysis of the discriminant $\Delta$ of this third-order polynomial,
$$\Delta=4\epsilon(1-27\epsilon k^2),$$
implies that
\begin{equation}
\label{discrim}
\left\{\begin{array}{lcl}
\Delta>0&\Longrightarrow&\mbox{there are three distinct real roots},\\[2ex]
\Delta=0&\Longrightarrow&\mbox{there is a multiple root and all of them are real},\\[2ex]
\Delta<0&\Longrightarrow&\mbox{there is one real root and two complex conjugate roots.}
\end{array}\right.
\end{equation}
This result clearly shows that $k$ can be seen as a bifurcation parameter which breaks the topological symmetry present in the case where we only have $\epsilon$ (RW metrics), because if we fix the sign of $\epsilon$, the sign of $\Delta$ still depends on the specific value of $k$. The details of this analysis can be widely extended using the qualitative theory of dynamical systems \cite{smale}, but this would deviate from our goals and it should be addressed in future work. The key point we learn from this analysis is that even fixing $sign(\epsilon)$, the value of $k$ is important for the determination of the domain of $\sigma(\chi)$, which directly interferes in the spatial features of the metric.

Finally, making the coordinate transformation given by $r=\sigma(\chi)$, the line element\ (\ref{fried}) becomes

\begin{equation}
\label{fried_r_pi}
ds^2=dt^2-a^2(t)\left(\fracc{dr^2}{1-\epsilon r^2-\frac{2k}{r}}+r^2d\Omega^2\right).
\end{equation}
One can see that the time evolution of this metric is still given by the usual Friedmann equations\ (\ref{gr_fried_solved1}) and\ (\ref{gr_fried_solved2}). In the conformal time $d\eta=dt/a(t)$, we get

\begin{equation}
\label{fried_r_pi_conf}
ds^2=a^2(\eta)\left(d\eta^2-\fracc{dr^2}{1-\epsilon r^2-\frac{2k}{r}}-r^2d\Omega^2\right).
\end{equation}

This line element leads us to comprehend the physical meaning of the arbitrary parameter $k$: it is closely related to the scale of homogeneity of the Universe; if $2k/r\ll1$ this term can be dropped out in\ (\ref{fried_r_pi}) and, hence, the Friedmann model is completely recovered in this regime. On the other hand, if $2k/r$ is not negligible, the 3-space is non-trivial and the particle trajectories in this metric have special features that we shall discuss in the next sections. Note also that this solution is not conformally equivalent to Schwarzschild-de Sitter metric. In the particular case $\epsilon=0$, the 3-space corresponds to the well-known Schwarzschild lattice (see Ref. \cite{rindler} for more details). From this similitude, we were led to analyze the Killing vectors and the Petrov classification of the metric\ (\ref{fried_r_pi}): it has only 3 Killing vectors, corresponding to the isotropy of the spacetime and, therefore, it is classified as Petrov-type D.

\section{Quasi-Maxwellian equations and Friedmann-like metrics}

The main point of this section is to show how the Quasi-Maxwellian equations of gravity shall be treated as complementary to the Einstein equations when we consider the initial condition problem and, as an example, we use them to reproduce the solution expressed by Eq.\ (\ref{fried_r_pi}). It should also be remarked that the QM equations deal directly with observable quantities. They are written in terms of the kinematical objects (expansion, shear, vorticity and acceleration), the energy-momentum tensor components (energy density, pressures and heat flux) and the components of the Weyl tensor (electric and magnetic parts).

It is convenient to assume the cosmic observer $V^{\mu}=\delta^{\mu}_0$ as before. The energy momentum distribution is given by a fluid with anisotropic pressure $\pi_{\mu\nu}$. Using the metric given in Eq.\ (\ref{fried}), these assumptions lead to the following Quasi-Maxwellian equations (see details in Appendix\ [\ref{appendix}]):

\begin{subequations}
\label{qm_gfried}
\begin{eqnarray}
&&\dot\theta+\fracc{\theta^2}{3}=-\fracc{1}{2}(\rho+3p),\label{qm_gfried1}\\[2ex]
&&\dot\rho+(\rho+p)\,\theta=0,\label{qm_gfried2}\\[2ex]
&&E_{\mu\nu}=-\fracc{1}{2}\pi_{\mu\nu},\label{qm_gfried3}\\[2ex]
&&E^{\alpha}{}_{\mu;\alpha}=0.\label{qm_gfried4}
\end{eqnarray}
\end{subequations}
As we said before, Eqs.\ (\ref{qm_gfried1}) and (\ref{qm_gfried2}) correspond to the Friedmann equations. Eq.\ (\ref{qm_gfried3}) yields immediately the electric part of the Weyl tensor in terms of the anisotropic pressure. Eq.\ (\ref{qm_gfried4}) represents the compatibility condition of the Einstein equations with the (constant) spatial curvature given by Eq.\ (\ref{dif_sigma}). Note that this equation does not imply that the Weyl tensor is identically zero.

From a straightforward calculation, the electric part of Weyl tensor for the cosmic observer reads

\begin{equation}
\label{el_weyl}
[E^i{}_j]=E(t,\chi)
\left(
\begin{array}{ccc}
1&0&0\\
0&-\frac{1}{2}&0\\
0&0&-\frac{1}{2}
\end{array}
\right),
\end{equation}
where

\begin{equation}
\label{el_weyl_fried}
E(t,\chi)=-\fracc{1}{3a^2}\left(\fracc{\sigma''}{\sigma}-\fracc{\sigma'^2}{\sigma^2}+\fracc{1}{\sigma^2}\right).
\end{equation}

Substituting this expression into Eq.\ (\ref{qm_gfried4}), we are left with an extra constraint, which is given by

\begin{equation}
\label{vinc_el_weyl}
-\fracc{1}{3a^2}\left(\fracc{\sigma''}{\sigma}-\fracc{\sigma'^2}{\sigma^2}+\fracc{1}{\sigma^2}\right)=\fracc{h(t)}{\sigma^3},
\end{equation}
where $h(t)$ is an arbitrary function. Multiplying both sides of Eq.\ (\ref{vinc_el_weyl}) by $a^2\sigma^3$, it follows

\begin{equation}
\label{vinc_el_weyl2}
\sigma^2\sigma''-\sigma\sigma'^2+\sigma=-3a^2(t)h(t).
\end{equation}
The l.h.s. is a function only of $\chi$ and the r.h.s. is a function only of time. Thus, the algebraic equation for the time coordinate is trivially satisfied by fixing $h=-k/a^2$, where $k$ is a constant. To solve the equation for the spatial coordinate $\chi$, we use the spatial curvature equation (\ref{3_ricci}) in terms of $\sigma$ and rewrite Eq.\ (\ref{vinc_el_weyl2}) as follows

\begin{equation}
\label{vinc_el_weyl2_3ricci}
\fracc{\sigma''}{\sigma}+\epsilon-\fracc{k}{\sigma^3}=0.
\end{equation}

Multiplying both sides of this equation by $2\sigma\sigma'$ and integrating it, we get

\begin{equation}
\label{intvinc_el_weyl2_3ricci1}
\sigma'^2+\epsilon\sigma^2+\fracc{2k}{\sigma}=C_1.
\end{equation}

The direct integration of the spatial curvature equation is expressed by

\begin{equation}
\label{intvinc_el_weyl2_3ricci2}
\sigma'^2+\epsilon\sigma^2-1-\fracc{C_2}{\sigma}=0.
\end{equation}
The common solutions of Eqs.\ (\ref{intvinc_el_weyl2_3ricci1}) and (\ref{intvinc_el_weyl2_3ricci2}) are obtained by setting $C_1=1$ and $C_2=-2k$.

Finally, we make the coordinate transformation $r=\sigma(\chi)$ to rewrite the line element\ (\ref{fried}) as given by Eq.\ (\ref{fried_r_pi}). Notice that, in this formalism, we know from the very beginning that the solution has a non-vanishing electric part of the Weyl tensor, which is given by

\begin{equation}
\label{el_weyl_r}
[E^i{}_j]=\fracc{k}{a^2r^3}
\left(
\begin{array}{ccc}
-1&0&0\\
0&1/2&0\\
0&0&1/2
\end{array}
\right),
\end{equation}
even assuming constant spatial curvature. Eq.\ (\ref{el_weyl_r}) is very similar to the Newtonian tidal forces multiplied by a time dependent function. Therefore, we conclude that Friedmann equations modify the Weyl tensor due to its time dependence via scale factor, but the presence of the Weyl tensor does not change the Friedmann equations. The consequences of this fact will be discussed in what follows.

\section{Trajectories of test particles}\label{test_part}

Instead of integrating completely the geodesic equations of the metric\ (\ref{fried_r_pi}), which seems unnecessary at this moment, a qualitative analysis of the particle trajectories in this geometry can be performed. In fact, we only analyze the geodesic motion of test particles (timelike and nulllike) moving along the equatorial plane ($\theta=\pi/2$ and $\dot\theta=0$). This simplification allows us to study the behavior of the effective potential to which these particles are subjected and to compare their paths with other cases.

In this way, the simplified geodesic equations reduce to

\begin{eqnarray}
\label{geo_pi}
&&t''+\fracc{a\dot a}{A}r'^2+a\dot ar^2\phi'^2=0,\label{geo_pi1}\\[2ex]
&&t'^2-\fracc{a^2}{A}r'^2-a^2r^2\phi'^2=b,\label{geo_pi2}\\[2ex]
&&\phi''+2\fracc{\dot a}{a}t'\phi'+2\fracc{r'}{r}\phi'=0,\label{geo_pi3}
\end{eqnarray}
where we denoted $X'\equiv dX/d\tau$, $\tau$ is the affine parameter along the curve (which is the proper time for the timelike geodesics) and $b$ is equal to $1$ for timelike geodesics or $0$ for nulllike ones. We also denoted
$$A(r)\equiv1-\epsilon\, r^2-\fracc{2k}{r}.$$
Note that we do not fix the value of the spatial curvature in the analysis: $\epsilon$ remains an arbitrary constant.

First, we solve Eq.\ (\ref{geo_pi3}) and see that the angular momentum is a conserved quantity

\begin{equation}
\label{int_geo_pi_phi}
\phi'=\fracc{l}{a^2r^2},
\end{equation}
where $l$ is an integration constant. Then, substituting Eqs.\ (\ref{int_geo_pi_phi}) and (\ref{geo_pi2}) into Eq.\ (\ref{geo_pi1}) we get

\begin{equation}
\label{pre_int_geo_pi_t}
t''+\fracc{\dot a}{a}(t'^2-b)=0,
\end{equation}
which can be integrated, resulting in

\begin{equation}
\label{int_geo_pi_t}
(t'^2-b)=\fracc{E}{a^2},
\end{equation}
where $E>0$ is another integration constant which is, in general, associated to the total energy of the test particle. Substituting this equation in Eq.\ (\ref{geo_pi2}) yields

\begin{equation}
\label{int_geo_pi_r}
a^4r'^2=\left(E-\frac{l^2}{r^2}\right)A.
\end{equation}
This equation can be seen as the energy conservation equation of a particle moving in a one-dimensional effective potential. Note that this equation has a constant term $E$ (like a mechanical energy), a kinetic-like term $a^4r'^2$ and the remaining ones correspond to the effective potential that we denote by $V(r)$.

To compare Eq.\ (\ref{int_geo_pi_r}) with the effective potential obtained from the Schwarzschild metric \cite{adler}, we set the spatial curvature equal to zero ($\epsilon=0$). Therefore, Eq.\ (\ref{int_geo_pi_r}) becomes

\begin{equation}
\label{int_geo_pi_r_e0}
a^4r'^2=E-\fracc{2kE}{r}-\frac{l^2}{r^2}+\frac{2kl^2}{r^3}.
\end{equation}
In this case $V(r)$ is given by

$$V(r)=\fracc{2kE}{r}+\frac{l^2}{r^2}-\frac{2k\,l^2}{r^3}.$$
From now on, we are only considering the case of vanishing spatial curvature $\epsilon=0$. This potential possesses almost the same terms as provided by the Schwarzschild metric for a single particle moving along geodesics, except the Newtonian term ($1/r$) which has a positive sign. Therefore, for particles moving radially ($l=0$) or for large values of $r$, the effective gravitational potential is apparently repulsive. It should also be remarked that when $r=2k$ the right hand side of Eq.\ (\ref{int_geo_pi_r}) is identically zero for massive particles and light rays. It means that the total energy is equal to the effective potential and, contrary to the Schwarzschild geodesics, this is a turning point for all test bodies. Again, we are led to conclude that the region $r<2k$ is excluded from this manifold.

For the sake of comparison, we analyze qualitatively the Kepler problem in this solution rewriting Eq.\ (\ref{int_geo_pi_r}) in terms of the variable $u=1/r$, where we seek for the planetary orbits $u=u(\phi)$. However, the time dependence of Eq.\ (\ref{int_geo_pi_r}) makes the problem more complicated. Therefore, we consider an interval of cosmological time $(t_0,t_1)$ in which the scale factor remains almost constant, i.e., $a(t)\equiv a_0$ for $t\in(t_0,t_1)$. In other words, we assume that the cosmological evolution is very slow when compared to the period of revolution around the center of symmetry. We thus obtain

\begin{equation}
\label{int_geo_pi_u}
l^2\left(\fracc{du}{d\phi}\right)^2=(E-l^2u^2)(1-2ku).
\end{equation}
Differentiating this equation with respect to $\phi$ yields

\begin{equation}
\label{diff_geo_pi_u}
\fracc{d^2u}{d\phi^2}+u=3ku^2-\fracc{Ek}{l^2}.
\end{equation}
This differential equation is very similar to the one given by the Schwarzschild metric in response to the Kepler problem if we interpret $k$ as the effective mass of the gravitational source and $E$ as the total mass of the test particle \cite{adler}. However, the inhomogenous term of the equation has a negative sign. Particularly, considering a massive test particle ($b>0$) we can reproduce perihelion shifts and, for $b=0$, we reproduce deflection of light rays, both predicted by the Schwarzschild solution of GR. The specific values for the shift and deflection may be different and only the complete integration of Eq.\ (\ref{diff_geo_pi_u}) can provide these numbers for our case. Notwithstanding, we intend to analyze this specific case in a future work.

These results indicate the importance of the Weyl tensor in order to describe local gravitational effects that cannot be interpreted as produced by any local ``visible" matter distribution. In other words, the Weyl tensor contains the information of global conditions imposed upon the spacetime, which modify the local behavior of particles and light rays.

\section{Geodesic deviation of the cosmological fluid}

Among its attributes, the Quasi-Maxwellian representation of gravity has the quality of putting together the formalism of the electromagnetic interactions and a formal approach of GR. Nonetheless, some fundamental distinctions must be stressed. The empirical determination of an electromagnetic field, for instance, is made through the Lorentz force and a test particle in order to identify the presence of the electromagnetic field. The electromagnetic tensor, obtained from the integration of Maxwell equations, does not distinguish the contribution of the local charge and current distributions from boundary conditions. In GR the empirical identification of a gravitational field cannot be made using a single test particle since the Christoffel symbols can be set equal to zero by coordinate transformations. In order to empirically determine the properties of the gravitational field it is necessary to look for the geodetic deviation expressed in terms of the curvature tensor. This tensor explicitly separates the
contribution coming from the local distribution of the energy-momentum tensor, algebraically associated to the traces of the Riemann tensor, from the global contribution of boundary conditions represented by the Weyl tensor.

Indeed, the measurements of the gravitational field effects can only be done through the geodesic deviation equation which determines the rate of the relative acceleration between two infinitesimally nearby geodesics, namely

\begin{equation}
\label{geo_dev}
\fracc{d^2z^{\alpha}}{ds^2}=-R^{\alpha}{}_{\beta\mu\nu}V^{\beta}z^{\mu}V^{\nu},
\end{equation}
where $z^{\alpha}$ is the deviation vector and $V^{\mu}$ is the vector field tangent to the geodesic congruence.

The distortion produced by the Weyl tensor upon a given congruence of curves can only be detected by this equation substituting the Riemann tensor by its decomposition into irreducible parts: the Ricci tensor, scalar curvature and the Weyl tensor. Since we are dealing with the comoving frame of the cosmological fluid, we set $V^{\mu}=\delta^{\mu}_0$. Evaluating the right hand side of Eq.\ (\ref{geo_dev}), we get
$$\fracc{d^2z^{\alpha}}{ds^2} = \left(E^{\alpha}{}_{\mu} + \frac{1}{2}\pi^{\alpha}{}_{\mu}\right) z^{\mu}+\fracc{1}{6}(\rho+3p)h^{\alpha}{}_{\mu}z^{\mu}.$$
Now comes a remarkable result: according to Eq.\ (\ref{qm_gfried3}) the term inside the big brackets is identically zero for our solution. Therefore, the cosmological fluid does not measure any distortion caused by the presence of the anisotropic pressure. In other words, the distortion caused by the anisotropic pressure and the electric part of the Weyl tensor are compensated in such a way that the cosmological observers do not attribute any eventual modification of the spacetime to these quantities, enabling one to set $E_{\mu\nu}$ and $\pi_{\mu\nu}$ equal to zero by hand. However, this is not allowed if we want to understand correctly the gravitational field effects in the Universe using the empirical data as initial conditions. Besides, the presence of the anisotropic pressure may change dramatically the perturbed version of the theory and hence the large scale structure formation. This will be investigated in a forthcoming paper.

\section{Concluding Remarks}

We have seen that the Einstein equations do not contain in its dynamics all the information necessary to determine the curvature tensor from empirical data. From this point of view, it means that the initial condition problem in GR (or the Cauchy problem) should be revisited in order to get a more realistic description of the universe. In other words, the Einstein equations correspond to an open system rather than a closed totality representing the universe, because at any time new elements, for instance the Weyl tensor, can play a role and modify some features of the spacetime, as we have presented in this paper.

In particular, we have shown how the standard cosmological model sets the Weyl tensor equal to zero \textit{ab initio} and that this is not a consequence of the Friedmann equations. Therefore, we have developed a cosmological model with constant spatial curvature and non-zero Weyl tensor without spoiling the conventional time evolution of the universe.

\section{acknowledgements}
We would like to thank Prof. M. Novello for many clarifying developments on QM equations. The authors also acknowledge CNPq for the financial support.

\section{Appendix: Quasi-Maxwellian equations}\label{appendix}

We know that a Riemannian geometry satisfies the Bianchi identities. In particular, in the case of general relativity, the Bianchi identities together with the Einstein equations yield the Quasi-Maxwellian equations of gravity. These equations are easily obtained if we rewrite the Riemann tensor in terms of its traces and the Weyl tensor:

\begin{equation}
\label{decom_riem}
R_{\alpha\beta\mu\nu}= W_{\alpha\beta\mu\nu} + M_{\alpha\beta\mu\nu} - \fracc{1}{6} R g_{\alpha\beta\mu\nu},
\end{equation}
where the auxiliary tensors are
\begin{equation}
\nonumber
2M_{\alpha\beta\mu\nu} \doteq R_{\alpha\mu}g_{\beta\nu} + R_{\beta\nu}g_{\alpha\mu} - R_{\alpha\nu}g_{\beta\mu} - R_{\beta\mu}g_{\alpha\nu}
\end{equation}
and
\begin{equation}
\nonumber
g_{\alpha\beta\mu\nu}\doteq g_{\alpha\mu}g_{\beta\nu}-g_{\alpha\nu}g_{\beta\mu}.
\end{equation}

Therefore, assuming Einstein constant equal to $1$, the Bianchi identities become

\begin{equation}
\label{bianchi}
W^{\alpha\beta\mu\nu}{}_{;\nu}=-\fracc{1}{2}T^{\mu[\alpha;\beta]}+
\fracc{1}{6}g^{\mu[\alpha}T^{,\beta]},
\end{equation}
where the square brackets mean anti-symmetrization. We use Eq.\ (\ref{decom_riem}) to define the Weyl tensor

\begin{equation}
\nonumber
W_{\alpha\beta\mu\nu}\doteq R_{\alpha\beta\mu\nu} - M_{\alpha\beta\mu\nu} + \fracc{1}{6} R g_{\alpha\beta\mu\nu},
\end{equation}

The reason of this nomenclature is due to several analogies between the Quasi-Maxwellian and the Maxwell equations. However, this similarity cannot be applied to the dynamical equations because the QM equations are in fact highly non-linear and of higher order of differentiability in comparison to Maxwell's theory, leading to situations that never happen in the last case. Indeed, the similitude appears when we make the projection of the QM equations with respect to the vector field $V^{\alpha}$ and its orthogonal hypersurface. At this point, it is very useful to replace the Weyl tensor by its electric $E_{\alpha\beta}$ and magnetic $H_{\alpha\beta}$ parts:

\begin{equation}
\nonumber
\begin{array}{lcl}
E_{\alpha\beta}&\doteq&-W_{\alpha\mu\beta\nu}V^{\mu}V^{\nu},\\[2ex]
H_{\alpha\beta}&\doteq&-^{*}W_{\alpha\mu\beta\nu}V^{\mu}V^{\nu},
\end{array}
\end{equation}
where $^{*}W_{\alpha\mu\beta\nu}$ is the dual of Weyl tensor constructed with the skew-symmetric Levi-Civita tensor. In the same way as the Faraday tensor $F_{\mu\nu}$, we can rewrite the Weyl tensor in terms of the quantities defined above

\begin{equation}
\nonumber
W_{\alpha\beta}{}^{\mu\nu} = 2\, V_{[\alpha}E_{\beta]}{}^{[\mu}V^{\nu]} + \delta_{[\alpha}^{[\mu}E_{\beta]}^{\nu]} - \eta_{\alpha\beta\lambda\sigma}V^{\lambda}H^{\sigma[\mu}V^{\nu]}-\eta^{\mu\nu\lambda\sigma}V_{\lambda}H_{\sigma[\alpha}V_{\beta]}.
\end{equation}

In parallel, the covariant derivative of $V^{\mu}$ can be also decomposed into its irreducible parts:

\begin{equation}
\nonumber
V_{\mu;\nu} = \sigma_{\mu\nu} + \omega_{\mu\nu} + \fracc{1}{3}\theta h_{\mu\nu} + a_{\mu}V_{\nu},
\end{equation}
where $\theta\doteq V^{\mu}{}_{;\mu}$ is the expansion coefficient, $a^{\mu}\doteq V^{\mu}{}_{;\nu}V^{\nu}$ is the acceleration,
$$\sigma_{\mu\nu} \doteq \frac{1}{2} h_{\mu}{}^{\alpha} h_{\nu}{}^{\beta} V_{(\alpha;\beta)} - \frac{\theta}{3} h_{\mu\nu}$$
is the shear tensor and
$$\omega_{\mu\nu}\doteq\frac{1}{2} h_{\mu}{}^{\alpha} h_{\nu}{}^{\beta} V_{[\alpha;\beta]}$$
is the vorticity.

With this in mind, the four independent projections of the Bianchi identities

\begin{equation}
\label{proj_div_weyl}
\begin{array}{l}
W^{\alpha\beta\mu\nu}{}_{;\nu}V_{\beta}V_{\mu}h_{\alpha}{}^{\sigma},\\[1ex]
W^{\alpha\beta\mu\nu}{}_{;\nu}\eta^{\sigma\lambda}{}_{\alpha\beta}V_{\mu}V_{\lambda},\\[1ex]
W^{\alpha\beta\mu\nu}{}_{;\nu}h_{\mu}{}^{(\sigma}\eta^{\tau)\lambda}{}_{\alpha\beta}V_{\lambda},\\[1ex]
W^{\alpha\beta\mu\nu}{}_{;\nu}V_{\beta}h_{\mu(\tau}h_{\sigma)\alpha},
\end{array}
\end{equation}
lead to the following linearly independent equations

\begin{equation}
\label{quase_max1}
h^{\epsilon\alpha}h^{\lambda\gamma}E_{\alpha\lambda;\gamma} + \eta^{\epsilon}{}_{\beta\mu\nu}V^{\beta}H^{\nu\lambda}\sigma^{\mu}{}_{\lambda} + 3H^{\epsilon\nu}\omega_{\nu} = \fracc{1}{3}h^{\epsilon\alpha}\rho_{,\alpha} + \, \fracc{\theta}{3}q^{\epsilon} -\fracc{1}{2}(\sigma^{\epsilon}{}_{\nu} - 3\omega^{\epsilon}{}_{\nu})q^{\nu} + \fracc{1}{2}\pi^{\epsilon\mu}a_{\mu} + \fracc{1}{2}h^{\epsilon\alpha}\pi_{\alpha}{}^{\nu}{}_{;\nu};
\end{equation}

\begin{equation}
\label{quase_max2}
h^{\epsilon\alpha}h^{\lambda\gamma}H_{\alpha\lambda;\gamma} - \eta^{\epsilon}{}_{\beta\mu\nu}V^{\beta}E^{\nu\lambda}\sigma^{\mu}{}_{\lambda} - 3E^{\epsilon\nu}\omega_{\nu} = (\rho+p)\omega^{\epsilon} - \,\fracc{1}{2}\eta^{\epsilon\alpha\beta\lambda}V_{\lambda}q_{\alpha;\beta} + \fracc{1}{2}\eta^{\epsilon\alpha\beta\lambda}(\sigma_{\mu\beta} + \omega_{\mu\beta})\pi^{\mu}{}_{\alpha}V_{\lambda};
\end{equation}

\begin{equation}
\label{quase_max3}
\begin{array}{l}
h_{\mu}{}^{\epsilon}h_{\nu}{}^{\lambda}\dot H^{\mu\nu} + \theta H^{\epsilon\lambda} - \fracc{1}{2}H_{\nu}{}^{(\epsilon}h^{\lambda)}_{\mu}V^{\mu;\nu} -a_{\alpha}E_{\beta}{}^{(\lambda}\eta^{\epsilon)\gamma\alpha\beta}V_{\gamma}+\,\eta^{\lambda\nu\mu\gamma}\eta^{\epsilon\beta\tau\alpha} V_{\mu}V_{\tau} H_{\alpha\gamma} \theta_{\nu\beta} + \fracc{1}{2}E_{\beta}{}^{\mu}{}_{;\alpha} h_{\mu}^{(\epsilon}\eta^{\lambda)\gamma\alpha\beta}V_{\gamma} =\\[2ex]
- \fracc{3}{4}q^{(\epsilon}\omega^{\lambda)} + \fracc{1}{2}h^{\epsilon\lambda}q^{\mu}\omega_{\mu}
+\fracc{1}{4}\sigma_{\beta}{}^{(\epsilon}\eta^{\lambda)\alpha\beta\mu}V_{\mu}q_{\alpha} +\fracc{1}{4}h^{\nu(\epsilon}\eta^{\lambda)\alpha\beta\mu}V_{\mu}\pi_{\nu\alpha;\beta};
\end{array}
\end{equation}

\begin{equation}
\label{quase_max4}
\begin{array}{l}
h_{\mu}{}^{\epsilon}h_{\nu}{}^{\lambda}\dot E^{\mu\nu} + \theta E^{\epsilon\lambda} - \fracc{1}{2}E_{\nu}{}^{(\epsilon}h^{\lambda)}{}_{\mu}V^{\mu;\nu} +a_{\alpha}H_{\beta}{}^{(\lambda}\eta^{\epsilon)\gamma\alpha\beta}V_{\gamma} + \eta^{\lambda\nu\mu\gamma}\eta^{\epsilon\beta\tau\alpha}V_{\mu}V_{\tau} E_{\alpha\gamma} \theta_{\nu\beta} - \fracc{1}{2}H_{\beta}{}^{\mu}{}_{;\alpha} h_{\mu}^{(\epsilon}\eta^{\lambda)\gamma\alpha\beta} V_{\gamma} =\\[2ex] \fracc{1}{6}h^{\epsilon\lambda}(q^{\mu}{}_{;\mu} -q^{\mu}a_{\mu} -\pi^{\mu\nu}\sigma_{\mu\nu})
-\fracc{1}{2}(\rho+p)\sigma^{\epsilon\lambda}+\fracc{1}{2}q^{(\epsilon}a^{\lambda)}-\fracc{1}{4}h^{\mu(\epsilon}h^{\lambda)\alpha}q_{\mu;\alpha} + \fracc{1}{2}h_{\alpha}{}^{\epsilon}h_{\mu}{}^{\lambda}\dot\pi^{\alpha\mu} + \fracc{1}{4}\pi_{\beta}{}^{(\epsilon}\sigma^{\lambda)\beta}+\\[2ex]
- \fracc{1}{4}\pi_{\beta}{}^{(\epsilon}\omega^{\lambda)\beta} + \fracc{1}{6}\theta\pi^{\epsilon\lambda}.
\end{array}
\end{equation}
These are the Quasi-Maxwellian equations and it is clear the similitude to the Maxwell equations: the first pair corresponds to $\nabla\cdot\vec E$ and $\nabla\cdot\vec H$, while the last pair gives the time evolution of $\vec H$ and $\vec E$, respectively. To obtain a self-consistent system of equations we need to add the energy-momentum tensor conservation law $T^{\mu\nu}{}_{;\nu}=0$, which gives

\begin{equation}
\label{expl_proj_conserv_mom_eneg1}
\dot\rho + (\rho+p)\theta + \dot q^{\mu}V_{\mu} + q^{\alpha}{}_{;\alpha} - \pi^{\mu\nu}\sigma_{\mu\nu} = 0,
\end{equation}
and
\begin{equation}
\label{expl_proj_conserv_mom_eneg2}
(\rho+p)a_{\alpha} - p_{\mu}h^{\mu}{}_{\alpha} + \dot q_{\mu}h^{\mu}{}_{\alpha} + \theta q_{\alpha} + q^{\nu}\theta_{\alpha\nu} + q^{\nu}\omega_{\alpha\nu} + \pi_{\alpha}{}^{\nu}{}_{;\nu} + \pi^{\mu\nu}\sigma_{\mu\nu}V_{\alpha} = 0.
\end{equation}

The integrability condition
$$V^{\alpha}{}_{;\mu;\nu}-V^{\alpha}{}_{;\nu;\mu}=R^{\alpha}{}_{\beta\mu\nu}V^{\beta},$$
applied to the observer field we have chosen, can be translated into evolution equations plus constraints for the kinematical quantities. Thereby, the evolution equations are

\begin{equation}
\label{evol_quant_cine1}
\dot\theta + \fracc{\theta^2}{3} + 2(\sigma^2 + \omega^2) - a^{\alpha}{}_{;\alpha}= -\fracc{1}{2}(\rho+3p),
\end{equation}

\begin{equation}
\label{evol_quant_cine2}
\begin{array}{l}
h_{\alpha}{}^{\mu}h_{\beta}{}^{\nu}\dot\sigma_{\mu\nu} + \fracc{1}{3}h_{\alpha\beta}(a^{\lambda}{}_{;\lambda}-2\sigma^2-2\omega^2) + a_{\alpha}a_{\beta}-\fracc{1}{2}h_{\alpha}{}^{\mu}h_{\beta}{}^{\nu}(a_{\mu;\nu} + a_{\nu;\mu}) + \fracc{2}{3}\theta\sigma_{\alpha\beta} + \sigma_{\alpha\mu}\sigma^{\mu}{}_{\beta}+\omega_{\alpha\mu}\omega^{\mu}{}_{\beta} =\\[2ex]
 -E_{\alpha\beta} - \fracc{1}{2}\pi_{\alpha\beta},
\end{array}
\end{equation}

\begin{equation}
\label{evol_quant_cine3}
h_{\alpha}{}^{\mu}h_{\beta}{}^{\nu}\dot\omega_{\mu\nu} - \fracc{1}{2}h_{\alpha}{}^{\mu}h_{\beta}{}^{\nu}(a_{\mu;\nu} - a_{\nu;\mu})+\fracc{2}{3}\theta\omega_{\alpha\beta} -\sigma_{\beta\mu}\omega^{\mu}{}_{\alpha}+\sigma_{\alpha\mu}\omega^{\mu}{}_{\beta}=0,
\end{equation}
and the constraint equations are

\begin{equation}
\label{eq_vinc_quant_cine1}
\fracc{2}{3}\theta_{,\mu}h^{\mu}{}_{\lambda} - (\sigma^{\alpha}{}_{\gamma} + \omega^{\alpha}{}_{\gamma})_{;\alpha}h^{\gamma}{}_{\lambda} - a^{\nu}(\sigma_{\lambda\nu} + \omega_{\lambda\nu}) =-q_{\lambda},
\end{equation}

\begin{equation}
\label{eq_vinc_quant_cine2}
\omega^{\alpha}{}_{;\alpha}+2\omega^{\alpha}a_{\alpha}=0,
\end{equation}

\begin{equation}
\label{eq_vinc_quant_cine3}
H_{\tau\lambda} = -\fracc{1}{2} h_{(\tau}{}^{\epsilon} h_{\lambda)}{}^{\alpha} \eta_{\epsilon}{}^{\beta\gamma\nu} V_{\nu}(\sigma_{\alpha\beta} + \omega_{\alpha\beta})_{;\gamma} + a_{(\tau}\omega_{\lambda)}.
\end{equation}

For the sake of comparison, this set of equations is exactly the same one presented in Hawking's pioneer work \cite{hawking_66} and by Ellis in \cite{ellis} (except for conventions adopted). These equations have been extensively used to perform covariant and gauge-invariant perturbation theory in cosmology and, as we said before, they propagate the solutions of the Einstein equations defined only on a given Cauchy surface to the whole spacetime.

\end{document}